# AI and Worker Well-Being:
# Differential Impacts Across Generational Cohorts and Genders

Voraprapa Nakavachara[*]

Faculty of Economics, Chulalongkorn University

16 February 2026

This paper investigates the relationship between AI use and worker well-being outcomes—mental health, job enjoyment, and physical health and safety—using microdata from the OECD AI Surveys across seven countries. The results reveal that AI users are significantly more likely to report improvements across all three outcomes, with effects ranging from 7.9% to 19.7%. However, these results vary by generation and gender. Generation Y (1981–1996) exhibits positive associations across all well-being dimensions, while Generation X (1965–1980) reports improvements primarily in mental health and job enjoyment. In contrast, Generation Z (1997–2012) shows positive associations only for job enjoyment, which may reflect higher baseline familiarity with digital technologies and therefore smaller marginal effects on other well-being dimensions. Baby Boomers (born before 1965) experience limited benefits, as they may not find these tools as engaging or useful. Women report stronger mental health gains, whereas men report greater improvements in physical health. These findings suggest that AI's workplace impact is uneven and shaped by demographic factors, career stage, and the nature of workers' roles.

[*]Contact Email: Voraprapa.n@chula.ac.th
Acknowledgement: The author gratefully acknowledges the OECD for providing access to the OECD AI Surveys of Employers and Workers.
Ethical Statement: Ethics approval was obtained from Chulalongkorn University (COA No. 489/68).

## 1. Introduction

Artificial intelligence (AI) is rapidly transforming how tasks are performed, decisions are made, and jobs are structured across industries. While much of the existing research focuses on AI's economic effects—such as productivity gains, labor market shifts, and job displacement (Brynjolfsson et al., 2025; Frey & Osborne, 2017; Noy & Zhang, 2023)— its implications for workers' psychological and physical well-being remain comparatively underexplored and conceptually unresolved. Yet well-being outcomes—including mental health, job enjoyment, and physical health and safety—are critical not only for individual quality of life but also for long-term organizational performance and the sustainability of AI adoption.

Existing evidence on AI and worker well-being is limited and fragmented. Some studies document improvements in health and job quality through reduced physical strain and task automation (Giuntella et al., 2025), while others emphasize heightened stress, technostress, and burnout, particularly among workers with lower technological confidence or adaptive capacity (Kim & Lee, 2024). These mixed findings suggest that the well-being consequences of AI are theoretically ambiguous and may depend crucially on worker characteristics and institutional context. However, systematic evidence that jointly examines multiple dimensions of well-being and explicitly accounts for worker heterogeneity remains scarce.

This study addresses several gaps in the literature. First, while prior research has predominantly focused on productivity and employment outcomes, far fewer studies examine non-pecuniary outcomes such as worker well-being. Second, existing evidence is largely country-specific, with limited cross-country analysis. We contribute by providing cross-national micro-level evidence on three dimensions of worker well-being—mental health, job enjoyment, and physical health and safety. Third, the literature rarely examines whether the well-being effects of AI differ systematically across demographic groups, often implicitly assuming uniform impacts. Addressing these gaps is essential for understanding not only whether AI affects worker well-being, but for whom and under what conditions.

The need to address these gaps is both theoretical and empirical. From a theoretical perspective, established frameworks suggest that AI introduces countervailing mechanisms: it can enhance well-being by automating physically demanding or redundant tasks and augmenting workers' capabilities, while simultaneously undermining well-being through technostress, skill obsolescence concerns, and perceived job insecurity. Which of these mechanisms dominates—and how this balance varies across workers—remains an open empirical question. Empirically, identifying heterogeneous well-being effects is crucial for informing workplace practices, training policies, and institutional responses that aim to promote inclusive and sustainable AI adoption.

To investigate these issues, this paper examines the relationship between AI use and worker well-being using microdata from the OECD AI Surveys of Employers and Workers across 7 OECD countries. These countries were selected because they represent advanced economies with relatively high AI diffusion, comparable labor market institutions, and harmonized survey instruments, allowing for meaningful cross-country comparison while minimizing confounding institutional heterogeneity.

Focusing explicitly on well-being rather than productivity, we analyze three key outcomes: mental health, job enjoyment, and physical health and safety. We place particular emphasis on heterogeneity by generational cohort and gender. Generational differences may reflect variation in digital nativity, career stage, and adjustment costs, while gender differences may arise from occupational segregation, task exposure, and disparities in perceived technological competence.

This study contributes to the literature in three main ways. First, it shifts the focus of the AI–labor literature from economic performance to worker well-being, highlighting non-pecuniary consequences of AI adoption. Second, it provides cross-country micro-level evidence using harmonized OECD data, enabling systematic comparison across institutional contexts. Third, it demonstrates that the relationship between AI use and well-being is heterogeneous, varying systematically across generational cohorts and between women and men, thereby shedding light on the distributional implications of AI adoption within the workforce.

This paper is organized as follows. Section 2 reviews the relevant literature and presents the conceptual framework. Section 3 describes the data and methodology. Section 4 reports the empirical results, including subgroup analyses by generation and gender. Section 5 concludes with a discussion of key findings and implications.

## 2. Literature Review and Conceptual Framework

Our conceptual framework draws on several established theories to explain why the relationship between AI use and worker well-being is theoretically ambiguous and potentially heterogeneous across workers. First, the Job Demands–Resources (JD–R) model highlights how AI may simultaneously increase job resources (e.g., task automation and efficiency gains) and job demands (e.g., technostress and learning pressure), leading to ambiguous effects on well-being. Second, task-based theories of technological change emphasize that AI affects workers primarily through the reallocation of tasks rather than complete job replacement, implying heterogeneous impacts depending on task composition. Third, insights from task–technology fit (TTF) theory underscore that workers differ in their ability and willingness to adapt to AI, suggesting variation in well-being outcomes across generations and genders.

### 2.1 Job Demands-Resources Model

The Job Demands–Resources (JD–R) model (Bakker & Demerouti, 2007; Demerouti et al., 2001) conceptualizes the workplace as a dual-process system in which employee well-being is shaped by the balance between job resources and job demands. Within this framework, AI can function as a job resource by automating routine and repetitive tasks, improving efficiency, and supporting decision-making. This process is associated with higher engagement, improved performance, and greater job satisfaction.

At the same time, AI may introduce new job demands, including heightened learning pressure, constant adaptation to evolving systems, and increased technological complexity. These demands can drain physical and mental energy, potentially triggering a health-impairment process characterized by stress, exhaustion, and burnout (Kim et al., 2024; Capozza et al., 2023). Consequently, the JD–R model suggests that the well-being consequences of AI adoption depend critically on how AI is implemented and integrated into work processes (i.e., mostly as job resources or mostly as job demands).

Empirical evidence shows mixed results. For example, longitudinal evidence from Germany finds that AI exposure is associated with improvements in self-rated health and health satisfaction (Giuntella et al., 2025). Nazareno & Schiff (2021) report that workers facing automation risk have lower stress. In manufacturing settings, AI has been shown to alleviate physical strain and contribute to better mental health outcomes by reducing workers' exposure to physically taxing labor (Wei & Li, 2022). As tedious or dangerous tasks are automated, workers often report higher levels of engagement and job satisfaction. In addition to task substitution, AI also functions as an augmentative tool that improves efficiency and reduces cognitive load. This frees up workers' time for more meaningful, complex, or strategic responsibilities, thereby enhancing their sense of autonomy and fulfillment (OECD, 2023b; Johnson et al., 2020). Overall, AI's potential to reallocate effort away from repetitive tasks toward higher-value work suggests a range of pathways through which it may positively influence mental health, physical well-being, and job enjoyment.

On the other hand, a growing body of research highlights how AI-related job anxiety is often fueled by fears of skill devaluation and the erosion of control over work processes. For instance, Méndez-Suárez et al. (2026) find that workers frequently perceive AI as a threat to their skill relevance and autonomy, leading to concerns about obsolescence and the need for constant reskilling. These perceived threats can contribute to heightened anxiety and depression (Johnson et al., 2020). AI adoption may also exacerbate stress and burnout, particularly among workers with lower technological confidence. Kim and Lee (2024) report that individuals less adept at adapting to new tools experience higher psychological strain when AI is introduced. Moreover, the OECD (2023a) identifies older and lower-skilled workers as especially vulnerable to the adverse effects of AI, citing lower digital readiness and greater resistance to change.

## 2.2 Task-Based Theories of Technological Change

Task-based theories of technological change propose that AI affects workers not by replacing entire occupations, but by reallocating tasks between human labor and machines. The impact of AI is therefore inherently heterogeneous, depending on the task composition of an occupation and whether specific tasks are more efficiently performed by humans or by machine capital (Acemoglu & Restrepo, 2018; Acemoglu & Restrepo, 2019a; Acemoglu & Restrepo, 2019b).

Within this framework, automation generates a displacement effect as machines take over tasks previously performed by humans, which may be counterbalanced by a reinstatement effect—the creation of new tasks in which human labor retains a comparative advantage. As a result, the content of workers' jobs evolves rather than disappearing entirely. Workers reallocate effort toward tasks that AI cannot fully automate or that are enhanced by AI through augmentation.

Task-based theories also suggest the possibility of generational differences in the impacts of AI. Workers from different generational cohorts can be unevenly distributed across occupations and tasks, reflecting both life-cycle factors and broader structural changes in the economy over time. Younger workers may be more likely to hold roles that emerged alongside digitalization, while older workers may be concentrated in occupations shaped by earlier technological regimes. As a result, exposure to automation and task augmentation—and the associated changes in job content—may vary across age cohorts, potentially leading to heterogeneous well-being outcomes.

Similarly, task-based perspectives suggest that gender differences may arise. Because men and women are unevenly distributed across occupations and tasks, their exposure to automation and task augmentation may differ. If AI disproportionately affects tasks in which men and women are differently concentrated, the resulting changes in work content, autonomy, and task intensity may translate into differences in well-being outcomes.

## 2.3 Task-Technology Fit Theory

Task–technology fit (TTF) theory suggests that technology improves outcomes not simply through adoption, but when its capabilities align with task requirements and user characteristics (Goodhue & Thompson; 1995). Performance gains occur when there is a strong fit between technological functionality and job demands, whereas poor alignment may generate strain rather than benefits.

Applied to AI, this framework suggests that well-being depends on how well AI systems match workers' tasks, skills, and psychological resources. A strong fit may enhance autonomy, competence, and job satisfaction, while a misfit may lead to technostress, anxiety, or work alienation if AI replaces meaningful task components or exceeds workers' adaptive capacity (Vendramin et al., 2021).

Generational and gender differences may shape task–technology fit. Older workers may face higher adjustment costs, while younger workers are often more technologically fluent and open to experimentation. Similarly, occupational sorting and differences in technological confidence may cause AI to align differently with men's and women's work. These variations suggest that the well-being effects of AI may differ across demographic groups.

There is growing empirical evidence suggesting both generational and gender differences in technology adoption. Workers born after 1980 tend to be more adaptable and open to adopting new technologies such as AI (Wei & Li, 2022). At the same time, AI adoption may pose psychological challenges for older workers. Méndez-Suárez et al. (2026) find that AI can threaten older workers' long-held workplace status or devalue tacit knowledge accumulated over decades, leading to feelings of vulnerability or displacement.

Along the gender dimension, Humlum and Vestergaard (2025) document a persistent gender gap in the use of generative AI tools such as ChatGPT, with women reporting lower familiarity and confidence. Similarly, Otis et al. (2024) find that this gap persists across regions, sectors, and occupations. Moreover, Lakomý et al. (2025) show that female workers—particularly those with lower education levels or employed in manual jobs—are more likely to hold negative attitudes toward digitalization. Such differences in adoption, confidence, and attitudes may shape how AI use translates into worker well-being outcomes.

### 2.4 Conceptual Framework

Drawing on the Job Demands–Resources (JD–R) model, task-based theories of technological change, and task–technology fit (TTF) theory, this study conceptualizes the relationship between AI use and worker well-being as both ambiguous and heterogeneous.

AI adoption alters workers' task environments and the balance between job resources and job demands. From a JD–R perspective, AI may function as a resource by automating routine tasks and improving efficiency, or as a demand by increasing learning pressure, technostress, and uncertainty. Task-based theories emphasize that AI primarily affects workers through task reallocation rather than job replacement, implying that well-being effects depend on task composition and whether AI complements or displaces workers' activities. TTF theory further highlights that these effects depend on how well AI systems align with workers' skills, tasks, and adaptive capacity.

Together, these perspectives suggest that AI may enhance well-being when it supports autonomy, skill use, and task complementarity, but may reduce well-being when it generates misalignment, stress, or perceived loss of control. They also imply systematic heterogeneity in AI's well-being effects across workers. Differences in task allocation, technological familiarity, and adaptation costs across generational cohorts and genders may shape how AI use translates into mental health, job enjoyment, and physical health outcomes. This paper empirically examines these theoretical predictions using cross-country worker-level data.

Specifically, this study addresses the following research questions:

RQ1: How is AI use related to workers' mental health, job enjoyment, and physical health and safety?

RQ2: Does this relationship differ across generational cohorts?

RQ3: Does this relationship differ by gender?

## 3. Data and Methodology

### 3.1 Data

We utilize secondary microdata from the OECD AI Surveys of Employers and Workers (OECD, 2023a), which are part of the OECD program on AI in Work, Innovation, Productivity and Skills (AI-WIPS). This study adopts a cross-sectional quantitative research design. The surveys were conducted between mid-January and mid-February 2022, covering workers and firms in the manufacturing and financial and insurance sectors across Austria, Canada, France, Germany, Ireland, the United Kingdom, and the United States.

The survey instruments were externally developed by the OECD in collaboration with Kantar Public and were designed to ensure cross-national comparability.[1] This author did not administer the survey and used the data with permission from the OECD.[2] For this paper, we focus exclusively on the workers' survey, which contains detailed information on workers' demographic characteristics, AI usage at work, and perceived impacts of AI on various job-related outcomes.

Our analysis examines the relationship between workers' use of AI and three dimensions of perceived worker well-being: mental health, job enjoyment, and physical health and safety. These outcomes are derived from the following survey questions:[3]

- "How do you think AI has changed your mental health and well-being in the workplace?"
- "How do you think AI has changed how much you enjoy your job?"
- "How do you think AI has changed your physical health and safety in the workplace?"

We assess whether workers who report using AI in their job are more likely to report improvements in these well-being outcomes compared to non-AI users.

To examine heterogeneity, we further analyze whether these associations differ by generational cohort and gender. Generational cohorts are defined as Generation Z (born 1997–2012), Generation Y (1981–1996), Generation X (1965–1980), and Baby Boomers or older (born 1964 or earlier).

Sample construction follows a transparent inclusion–exclusion process. The initial worker survey sample consists of 5,334 respondents. Because questions on AI use were asked only to workers employed in firms that had adopted AI, the analytical sample was reduced to 2,927 respondents. Ten respondents who did not report binary gender (male or female) were excluded for the purpose of gender-based analysis, resulting in a final analytical sample of 2,917 workers, which is used consistently across all regressions.

---

[1] The instruments underwent an extensive validation process, including pre-testing in Germany and the United States, with cognitive testing and debriefing to ensure clarity, consistency, and cross-country comparability.

[2] Ethical approval for this research was obtained from Chulalongkorn University (COA No. 489/68). In the original OECD survey, participation was entirely voluntary, and respondents were informed of their rights, including the right to withdraw consent at any time.

[3] Note that the specific wording asked to non-AI users is slightly different, as they do not personally use AI. They are asked: "How do you think AI has changed the mental health and well-being of workers in your company?", "How do you think AI has changed how much workers in your company enjoy their job?", and "How do you think AI has changed the physical health and safety of workers in your company?" For the purpose of this study, we combine AI users and non-AI users to conduct the empirical analyses.

Table 1 reports summary statistics for the full sample, while Table 2 presents descriptive statistics disaggregated by generational cohort and Table 3 by gender. These tables provide background information on AI use, perceived well-being outcomes, demographic characteristics, employment status, and sector (manufacturing vs. finance).

Table 1: Summary Statistics (All Observations)

| Variable | Obs | Mean | Std. dev. | Min | Max |
|---|---|---|---|---|---|
| Improved Mental Health | 2,917 | 0.5207 | 0.4997 | 0 | 1 |
| Improved Job Enjoyment | 2,917 | 0.5876 | 0.4924 | 0 | 1 |
| Improved Physical Health | 2,917 | 0.5499 | 0.4976 | 0 | 1 |
| AI User | 2,917 | 0.7316 | 0.4432 | 0 | 1 |
| Bachelor or Higher | 2,917 | 0.6239 | 0.4845 | 0 | 1 |
| Male | 2,917 | 0.5927 | 0.4914 | 0 | 1 |
| Full-time | 2,917 | 0.9129 | 0.2820 | 0 | 1 |
| Job Satisfied | 2,917 | 0.3432 | 0.4748 | 0 | 1 |
| Manufacturing | 2,917 | 0.4457 | 0.4971 | 0 | 1 |

| Variable | | Freq. | Percent | Cum. |
|---|---|---|---|---|
| Generation | Gen Z (1997-2012) | 348 | 11.93 | 11.93 |
| | Gen Y (1981-1996) | 1,408 | 48.27 | 60.2 |
| | Gen X (1965-1980) | 952 | 32.64 | 92.84 |
| | Baby Boomers (1964 & Older) | 209 | 7.16 | 100 |
| Country | Austria | 345 | 11.83 | 11.83 |
| | Canada | 453 | 15.53 | 27.36 |
| | Germany | 430 | 14.74 | 42.1 |
| | Ireland | 290 | 9.94 | 52.04 |
| | UK | 419 | 14.36 | 66.4 |
| | USA | 508 | 17.42 | 83.82 |
| | France | 472 | 16.18 | 100 |

Table 2: Summary Statistics (by Generation)

Panel A: Gen Z (1997-2012)

| Variable | Obs | Mean | Std. dev. | Min | Max |
|---|---|---|---|---|---|
| Improved Mental Health | 348 | 0.5661 | 0.4963 | 0 | 1 |
| Improved Job Enjoyment | 348 | 0.6322 | 0.4829 | 0 | 1 |
| Improved Physical Health | 348 | 0.5891 | 0.4927 | 0 | 1 |
| AI User | 348 | 0.8161 | 0.3880 | 0 | 1 |
| Bachelor or Higher | 348 | 0.6121 | 0.4880 | 0 | 1 |
| Male | 348 | 0.6063 | 0.4893 | 0 | 1 |
| Full-time | 348 | 0.8477 | 0.3598 | 0 | 1 |
| Job Satisfied | 348 | 0.4655 | 0.4995 | 0 | 1 |
| Manufacturing | 348 | 0.3851 | 0.4873 | 0 | 1 |

Panel B: Gen Y (1981-1996)

| Variable | Obs | Mean | Std. dev. | Min | Max |
|---|---|---|---|---|---|
| Improved Mental Health | 1,408 | 0.5653 | 0.4959 | 0 | 1 |
| Improved Job Enjoyment | 1,408 | 0.6477 | 0.4778 | 0 | 1 |
| Improved Physical Health | 1,408 | 0.6044 | 0.4892 | 0 | 1 |
| AI User | 1,408 | 0.7834 | 0.4121 | 0 | 1 |
| Bachelor or Higher | 1,408 | 0.6903 | 0.4625 | 0 | 1 |
| Male | 1,408 | 0.5952 | 0.4910 | 0 | 1 |
| Full-time | 1,408 | 0.9219 | 0.2685 | 0 | 1 |
| Job Satisfied | 1,408 | 0.3594 | 0.4800 | 0 | 1 |
| Manufacturing | 1,408 | 0.4332 | 0.4957 | 0 | 1 |

Panel C: Gen X (1965-1980)

| Variable | Obs | Mean | Std. dev. | Min | Max |
|---|---|---|---|---|---|
| Improved Mental Health | 952 | 0.4632 | 0.4989 | 0 | 1 |
| Improved Job Enjoyment | 952 | 0.5042 | 0.5002 | 0 | 1 |
| Improved Physical Health | 952 | 0.4779 | 0.4998 | 0 | 1 |
| AI User | 952 | 0.6733 | 0.4692 | 0 | 1 |
| Bachelor or Higher | 952 | 0.5651 | 0.4960 | 0 | 1 |
| Male | 952 | 0.5830 | 0.4933 | 0 | 1 |
| Full-time | 952 | 0.9317 | 0.2524 | 0 | 1 |
| Job Satisfied | 952 | 0.2784 | 0.4484 | 0 | 1 |
| Manufacturing | 952 | 0.4842 | 0.5000 | 0 | 1 |

Panel D: Baby Boomers (1964 & Older)

| Variable | Obs | Mean | Std. dev. | Min | Max |
|---|---|---|---|---|---|
| Improved Mental Health | 209 | 0.4067 | 0.4924 | 0 | 1 |
| Improved Job Enjoyment | 209 | 0.4880 | 0.5011 | 0 | 1 |
| Improved Physical Health | 209 | 0.4450 | 0.4982 | 0 | 1 |
| AI User | 209 | 0.5072 | 0.5011 | 0 | 1 |
| Bachelor or Higher | 209 | 0.4641 | 0.4999 | 0 | 1 |
| Male | 209 | 0.5981 | 0.4915 | 0 | 1 |
| Full-time | 209 | 0.8756 | 0.3308 | 0 | 1 |
| Job Satisfied | 209 | 0.3254 | 0.4696 | 0 | 1 |
| Manufacturing | 209 | 0.4545 | 0.4991 | 0 | 1 |

Table 3: Summary Statistics (by Gender)

Panel A: Male Observations

| Variable | Obs | Mean | Std. dev. | Min | Max |
|---|---|---|---|---|---|
| Improved Mental Health | 1,729 | 0.5720 | 0.4949 | 0 | 1 |
| Improved Job Enjoyment | 1,729 | 0.6281 | 0.4834 | 0 | 1 |
| Improved Physical Health | 1,729 | 0.6084 | 0.4882 | 0 | 1 |
| AI User | 1,729 | 0.7785 | 0.4154 | 0 | 1 |
| Bachelor or Higher | 1,729 | 0.6466 | 0.4782 | 0 | 1 |
| Full-time | 1,729 | 0.9508 | 0.2163 | 0 | 1 |
| Job Satisfied | 1,729 | 0.3846 | 0.4866 | 0 | 1 |
| Manufacturing | 1,729 | 0.4980 | 0.5001 | 0 | 1 |

Panel B: Female Observations

| Variable | Obs | Mean | Std. dev. | Min | Max |
|---|---|---|---|---|---|
| Improved Mental Health | 1,188 | 0.4461 | 0.4973 | 0 | 1 |
| Improved Job Enjoyment | 1,188 | 0.5286 | 0.4994 | 0 | 1 |
| Improved Physical Health | 1,188 | 0.4646 | 0.4990 | 0 | 1 |
| AI User | 1,188 | 0.6633 | 0.4728 | 0 | 1 |
| Bachelor or Higher | 1,188 | 0.5909 | 0.4919 | 0 | 1 |
| Full-time | 1,188 | 0.8577 | 0.3495 | 0 | 1 |
| Job Satisfied | 1,188 | 0.2828 | 0.4506 | 0 | 1 |
| Manufacturing | 1,188 | 0.3695 | 0.4829 | 0 | 1 |

### 3.2 Methodology

To evaluate the association between AI use and worker well-being, we estimate the following probit regression model:

$$\Pr\,(y_i = 1) \; = \mathrm{g}(\beta_0 + \beta_1 ai_{user\,i} + \beta_2' x_i + manuf_i + \theta_g + \lambda_s + \gamma_c) \qquad (1)$$

where $y_i$ is a binary dependent variable that equals 1 if worker $i$ reports an improvement in (i) mental health, (ii) job enjoyment, or (iii) physical health and safety associated with AI use in the workplace, and 0 otherwise.

For each of these outcomes, the original survey responses include six options: "improved a lot," "improved a little," "worsened a little," "worsened a lot," "no effect," and "don't know." We recode $y_i = 1$ if the respondent answered "improved a lot" or "improved a little," and $y_i = 0$ otherwise. Separate regressions are estimated for each well-being dimension.

The key independent variable is $ai_user_i$, a dummy variable equal to 1 if the worker reports using AI in their job and 0 otherwise. Based on the existing literature, the expected sign of $\beta_1$ is theoretically ambiguous. For example, Giuntella, König, and Stella (2025) find no significant deterioration in workers' mental health due to AI exposure in Germany and even suggest improvements through reduced physical strain. In contrast, Kim and Lee (2024) report that AI adoption can increase stress and burnout, particularly among workers with low self-efficacy in adapting to new technologies. These mixed findings motivate the need for cross-country empirical evidence, which this study provides using OECD microdata.

The vector $x_i$ includes individual-level control variables that are motivated by the conceptual framework. Education is a dummy variable equal to 1 if the worker holds a bachelor's degree or higher and captures differences in skill levels and adaptive capacity. Gender (equal to 1 for male) and employment type (equal to 1 for full-time employment) account for variation in task allocation, job stability, and exposure to AI-related changes in work organization.

To account for baseline differences in workers' overall work experiences that may be correlated with both AI use and perceived well-being, we include a control for overall job satisfaction. This variable is derived from the question "How satisfied are you with your job?" and is coded as 1 if the worker reports being "very satisfied" or "somewhat satisfied," and 0 otherwise.

$Manuf_i$ is a dummy variable equal to 1 if the worker is employed in the manufacturing sector (as opposed to finance and insurance). $\theta_g$ represents generational cohort fixed effects—Generation Z (1997–2012), Generation Y (1981–1996), Generation X (1965–1980), and Baby Boomers or older (born 1964 or earlier). $\lambda_s$ denotes firm-size fixed effects, and $\gamma_c$ captures country fixed effects to control for cross-country institutional differences. Standard errors are clustered at the country level.

Futhermore, we re-estimate Equation (1) separately by generational cohort and by gender. This subgroup analysis allows us to assess whether the association between AI use and perceived well-being differs systematically across age groups and between male and female workers.

## 4. Results

### 4.1 Main Results

Table 4 reports the main regression results (marginal effects obtained from the probit model) for the three outcomes: improved mental health, improved job enjoyment, and improved physical health and safety. Across all specifications and outcomes, the coefficient on AI User is positive, sizable, and statistically significant. Specifically, AI users are 12.4 percentage points more likely to report improved mental health, 19.7 percentage points more likely to report improved job enjoyment, and 9.0 percentage points more likely to report improved physical health relative to non-users.

Table 4: Main Regression Results (Probit Model: Marginal Effects)

| VARIABLES | (1)<br>Improved Mental Health | (2)<br>Improved Mental Health | (3)<br>Improved Enjoyment | (4)<br>Improved Enjoyment | (5)<br>Improved Physical Health | (6)<br>Improved Physical Health |
|---|---|---|---|---|---|---|
| ai user | 0.124*** | 0.109*** | 0.197*** | 0.183*** | 0.0900*** | 0.0785*** |
| | (0.0720 - 0.176) | (0.0519 - 0.167) | (0.143 - 0.252) | (0.129 - 0.238) | (0.0499 - 0.130) | (0.0381 - 0.119) |
| bachelor or higher | 0.0978*** | 0.0749*** | 0.142*** | 0.122*** | 0.123*** | 0.105*** |
| | (0.0464 - 0.149) | (0.0224 - 0.128) | (0.0965 - 0.188) | (0.0820 - 0.161) | (0.0772 - 0.169) | (0.0656 - 0.144) |
| male | 0.0933*** | 0.0752** | 0.0549* | 0.0390 | 0.0894*** | 0.0748*** |
| | (0.0292 - 0.157) | (0.0135 - 0.137) | (-0.00746 - 0.117) | (-0.0232 - 0.101) | (0.0406 - 0.138) | (0.0286 - 0.121) |
| fulltime | 0.0350 | 0.0256 | 0.0440 | 0.0351 | 0.0438 | 0.0373 |
| | (-0.0271 - 0.0970) | (-0.0339 - 0.0852) | (-0.0222 - 0.110) | (-0.0261 - 0.0963) | (-0.0190 - 0.107) | (-0.0259 - 0.101) |
| job_satisfied | | 0.191*** | | 0.179*** | | 0.155*** |
| | | (0.164 - 0.217) | | (0.143 - 0.214) | | (0.129 - 0.182) |
| generation = 2, Gen Y (1981-1996) | 0.0150 | 0.0341 | 0.0264 | 0.0449 | 0.0226 | 0.0382 |
| | (-0.0577 - 0.0877) | (-0.0317 - 0.0999) | (-0.0462 - 0.0990) | (-0.0196 - 0.109) | (-0.0548 - 0.1000) | (-0.0315 - 0.108) |
| generation = 3, Gen X (1965-1980) | -0.0445 | -0.0190 | -0.0603 | -0.0351 | -0.0703 | -0.0490 |
| | (-0.133 - 0.0438) | (-0.0980 - 0.0601) | (-0.148 - 0.0278) | (-0.120 - 0.0497) | (-0.161 - 0.0207) | (-0.136 - 0.0379) |
| generation = 4, Baby Boomers (1964 & Older) | -0.0648 | -0.0512 | -0.0164 | -0.00313 | -0.0581 | -0.0466 |
| | (-0.160 - 0.0301) | (-0.139 - 0.0363) | (-0.169 - 0.136) | (-0.156 - 0.150) | (-0.138 - 0.0223) | (-0.125 - 0.0314) |
| manufacturing | 0.0171 | 0.0313 | 0.0210 | 0.0343 | 0.154*** | 0.166*** |
| | (-0.0338 - 0.0680) | (-0.0203 - 0.0829) | (-0.0272 - 0.0692) | (-0.0106 - 0.0792) | (0.107 - 0.202) | (0.119 - 0.213) |
| Observations | 2,917 | 2,917 | 2,917 | 2,917 | 2,917 | 2,917 |
| Company Size | YES | YES | YES | YES | YES | YES |
| Country | YES | YES | YES | YES | YES | YES |
| Obs | All | All | All | All | All | All |

ci in parentheses

*** p<0.01, ** p<0.05, * p<0.1

When overall job satisfaction is included as an additional control variable, the coefficients on AI use remain positive, statistically significant, and relatively stable in magnitude. The effects become 10.9 percentage points for improved mental health, 18.3 percentage points for improved job enjoyment, and 7.85 percentage points for improved physical health.

Job satisfaction itself emerges as one of the strongest predictors of perceived well-being changes related to AI. Workers who report being satisfied with their jobs are 19.1% more likely to report improved mental health, 17.9% more likely to report improved job enjoyment, and 15.5% more likely to report improved physical health across the three model specifications. These results suggest that although job satisfaction captures important underlying attitudes toward work, AI usage continues to exhibit a robust and independently positive association with workers' perceptions of mental, physical, and enjoyment-related outcomes.

For all subsequent subgroup analyses—those estimated separately by generation and by gender—we rely on the full specification that includes job satisfaction as part of the independent variables, given its strong predictive power and its role in capturing underlying attitudes toward work.

Education is also strongly associated with positive perceptions of AI. Workers with at least a bachelor's degree are between 7.49% more likely to report improved mental health, 12.2% more likely to report improved enjoyment, and 10.5% more likely to report improved physical health.

Gender differences appear in some models. Male workers are significantly more likely than female workers to report improvements in mental and physical health. Differences across industries are less pronounced. The manufacturing dummy is not statistically significant for mental health and job enjoyment but shows a strong and positive association with improved physical health. One plausible explanation could be that there are more tangible safety benefits of AI in physical production environments.

Overall, the main results provide strong evidence that AI use is positively associated with workers' perceptions of mental well-being, job enjoyment, and physical health, even after controlling for demographic characteristics, job characteristics, job satisfaction, firm size, and country fixed effects.

### 4.2 Results by Generation

Table 5 presents the regression results by generational cohort—Generation Z (1997–2012), Generation Y (1981–1996), Generation X (1965–1980), and Baby Boomers or older (born 1964 or earlier). Clear generational differences emerge in how workers well-being is associated with the use of AI.

Table 5: Regression Results by Generation (Probit Model: Marginal Effects)

| | (1) | (2) | (3) | (4) | (5) | (6) | (7) | (8) | (9) | (10) | (11) | (12) |
|---|---|---|---|---|---|---|---|---|---|---|---|---|
| VARIABLES | Improved Mental Health | Improved Mental Health | Improved Mental Health | Improved Mental Health | Improved Enjoyment | Improved Enjoyment | Improved Enjoyment | Improved Enjoyment | Improved Physical Health | Improved Physical Health | Improved Physical Health | Improved Physical Health |
| ai user | -0.0115 | 0.164*** | 0.0658* | 0.108** | 0.145*** | 0.192*** | 0.196*** | 0.110 | -0.0108 | 0.123*** | 0.0438 | 0.0528 |
| | (-0.136 - 0.113) | (0.0744 - 0.253) | (-0.00290 - 0.135) | (0.0107 - 0.205) | (0.0536 - 0.237) | (0.124 - 0.261) | (0.127 - 0.264) | (-0.0327 - 0.253) | (-0.150 - 0.129) | (0.0731 - 0.174) | (-0.0316 - 0.119) | (-0.0539 - 0.160) |
| bachelor or higher | 0.0540 | 0.0527** | 0.132*** | -0.0758 | 0.118*** | 0.128*** | 0.120*** | 0.0876 | 0.0331 | 0.0863*** | 0.159*** | 0.0227 |
| | (-0.0423 - 0.150) | (0.00851 - 0.0969) | (0.0641 - 0.199) | (-0.229 - 0.0774) | (0.0610 - 0.175) | (0.0851 - 0.171) | (0.0321 - 0.208) | (-0.0245 - 0.200) | (-0.0224 - 0.0885) | (0.0259 - 0.147) | (0.0960 - 0.222) | (-0.101 - 0.146) |
| male | 0.0332 | 0.109** | 0.0412 | 0.0759 | 0.0160 | 0.0714* | 0.0375 | -0.0372 | 0.143*** | 0.100*** | 0.0176 | 0.0357 |
| | (-0.0703 - 0.137) | (0.0201 - 0.197) | (-0.0441 - 0.127) | (-0.0212 - 0.173) | (-0.0713 - 0.103) | (-0.00338 - 0.146) | (-0.0538 - 0.129) | (-0.126 - 0.0517) | (0.0529 - 0.233) | (0.0270 - 0.173) | (-0.0466 - 0.0818) | (-0.0299 - 0.101) |
| fulltime | -0.0120 | 0.00904 | 0.0854 | -0.0380 | -0.0104 | 0.0183 | 0.0894 | -0.0103 | 0.00141 | 0.0747** | 0.0610 | -0.151 |
| | (-0.150 - 0.126) | (-0.0751 - 0.0932) | (-0.0685 - 0.239) | (-0.210 - 0.134) | (-0.190 - 0.169) | (-0.0256 - 0.0622) | (-0.0682 - 0.247) | (-0.160 - 0.140) | (-0.150 - 0.152) | (0.00619 - 0.143) | (-0.0261 - 0.148) | (-0.352 - 0.0495) |
| job_satisfied | 0.0973* | 0.235*** | 0.162*** | 0.143** | 0.0409 | 0.198*** | 0.200*** | 0.162*** | 0.0440 | 0.173*** | 0.168*** | 0.115 |
| | (-0.00861 - 0.203) | (0.184 - 0.285) | (0.120 - 0.205) | (0.0253 - 0.261) | (-0.0516 - 0.133) | (0.164 - 0.232) | (0.150 - 0.251) | (0.0524 - 0.272) | (-0.0758 - 0.164) | (0.130 - 0.216) | (0.105 - 0.230) | (-0.0456 - 0.275) |
| manufacturing | -0.0533 | 0.00970 | 0.0794** | 0.129 | 0.0161 | 0.0228 | 0.0548 | 0.0503 | 0.0422 | 0.117*** | 0.255*** | 0.278*** |
| | (-0.163 - 0.0560) | (-0.0576 - 0.0770) | (0.00163 - 0.157) | (-0.0329 - 0.290) | (-0.0483 - 0.0806) | (-0.0234 - 0.0689) | (-0.0188 - 0.128) | (-0.112 - 0.212) | (-0.123 - 0.207) | (0.0500 - 0.185) | (0.213 - 0.296) | (0.125 - 0.431) |
| Observations | 348 | 1,408 | 952 | 205 | 348 | 1,408 | 952 | 209 | 348 | 1,408 | 952 | 205 |
| Company Size | YES | YES | YES | YES | YES | YES | YES | YES | YES | YES | YES | YES |
| Country | YES | YES | YES | YES | YES | YES | YES | YES | YES | YES | YES | YES |
| Obs | Gen Z | Gen Y | Gen X | Baby Boomers | Gen Z | Gen Y | Gen X | Baby Boomers | Gen Z | Gen Y | Gen X | Baby Boomers |

ci in parentheses

*** p<0.01, ** p<0.05, * p<0.1

For Generation Z, AI use is not significantly associated with improved mental health or improved physical health. The only consistent and statistically significant outcome is improved job enjoyment, with magnitude of 14.5%. One plausible explanation could be that Gen Z workers—many of whom are first-jobbers, junior staff, or in early career entry-level positions—are already digital natives, with 81.61% reporting AI use. Their familiarity with technology may limit the incremental impact of AI on mental or physical strain, while still enhancing job enjoyment through novelty or task efficiency.

For Generation Y, the associations between AI use and well-being are the strongest, most consistent, and statistically significant across all three well-being outcomes. AI users in this cohort are 16.4% more likely to report improved mental health, 19.2% more likely to report improved job enjoyment, and 12.3% more likely to report improved physical health. A plausible explanation could be that Gen Y workers, many of whom occupy roles such as managers, supervisors, mid-level professionals, and heads of unit, may be in positions where AI tools tangibly reshape workflows, reduce routine burdens, and improve both psychological and physical aspects of their work, as well as contribute directly to higher job enjoyment.

Among Generation X workers, AI use is also associated with improved mental health and job enjoyment, but not with physical health. Specifically, AI use is associated with a 6.58% increase in the likelihood of reporting improved mental health and a 19.6% increase in improved job enjoyment. One tentative explanation is that many Generation X workers occupy senior supervisory, managerial, or specialist roles, where AI tools may support decision-making and efficiency, thereby enhancing emotional and job-related satisfaction. The absence of a significant association with physical health may reflect differences in task composition or potentially lower adaptability relative to younger cohorts, although this interpretation should be viewed with caution.

For Baby Boomers or older workers, the estimated associations between AI use and well-being are largely statistically insignificant for job enjoyment and physical health. There is some evidence of a positive association with mental health, with an estimated effect of 10.8%. One possible interpretation is that older workers—who are often in late-career, advisory, or senior leadership roles—may experience more limited changes in day-to-day task content following AI adoption, which could explain the weaker associations with job enjoyment. Similarly, the absence of significant associations with physical health may be related to differences in task composition or adaptation patterns across age groups, rather than to direct effects of AI use on physical well-being.

Taken together, the results indicate that the positive association between AI use and worker well-being is most pronounced for Generation Y, followed by Generation X. For Generation Z, the association is primarily observed for job enjoyment, while for Baby Boomers, positive associations are evident mainly for mental health, with no statistically significant associations for job enjoyment or physical health. Overall, these generational patterns suggest that differences across career stages, technological familiarity, and job roles may be relevant factors in understanding how AI use is associated with mental, physical, and job-related well-being.

### 4.3 Results by Gender

Table 6 presents the regression estimates by gender. The results reveal that AI use is positively associated with improved mental health, job enjoyment, and physical health, although the magnitudes of these associations differ by gender.

For male workers, AI use is associated with higher likelihoods of reporting improved mental health (9.07%), job enjoyment (17.8%), and physical health (8.36%). For female workers, AI use is also positively associated with improved mental health (11.8%), job enjoyment (18.6%), and physical health (7.05%). While associations are positive for both groups, the magnitudes suggest relatively stronger mental health associations for women and stronger physical health associations for men, with broadly similar associations for job enjoyment.

Across all specifications, education and job satisfaction remain strong correlates of well-being outcomes for both men and women. Holding a bachelor's degree or higher is associated with a 6.89–12.7% higher likelihood of reporting improved well-being, while job satisfaction is consistently associated with improvements of approximately 16.1–23.1% across mental health, job enjoyment, and physical health.

Taken together, the gender-specific results indicate that AI use is positively associated with multiple dimensions of worker well-being for both men and women. Differences in the relative magnitudes of these associations across genders may reflect variation in occupational roles, task environments, or the types of AI applications encountered at work, although the data do not allow these mechanisms to be directly tested. Overall, the findings suggest that the relationship between AI use and well-being is not uniform across genders and may be shaped by existing patterns of work organization and task allocation.

### 4.4 Discussion

Overall, the empirical evidence shows a clear and robust positive association between AI use and workers' perceived well-being across mental health, job enjoyment, and physical health. These patterns hold across all model specifications.

The subgroup analyses reveal substantial differences in these relationships. Across generational cohorts, Generation Y exhibits the strongest and most comprehensive associations, with positive relationships across all three well-being dimensions. Generation X also shows positive associations, particularly for mental health and job enjoyment, though with smaller magnitudes. Generation Z—despite having the highest rates of AI use—exhibits positive associations mainly for job enjoyment, while associations with mental and physical health are weaker. For Baby Boomers, positive associations are limited to mental health, with no significant relationships observed for job enjoyment or physical health.

Table 6: Regression Results by Gender (Probit Model: Marginal Effects)

| VARIABLES | (1)<br>Improved Mental Health | (2)<br>Improved Mental Health | (3)<br>Improved Enjoyment | (4)<br>Improved Enjoyment | (5)<br>Improved Physical Health | (6)<br>Improved Physical Health |
|---|---|---|---|---|---|---|
| ai user | 0.0907*** | 0.118*** | 0.178*** | 0.186*** | 0.0836*** | 0.0705* |
| | (0.0402 - 0.141) | (0.0294 - 0.206) | (0.103 - 0.254) | (0.123 - 0.248) | (0.0268 - 0.140) | (-0.00657 - 0.147) |
| bachelor or higher | 0.0777** | 0.0689** | 0.115*** | 0.127*** | 0.119*** | 0.0733*** |
| | (0.0126 - 0.143) | (0.00706 - 0.131) | (0.0676 - 0.162) | (0.0659 - 0.188) | (0.0659 - 0.173) | (0.0272 - 0.119) |
| fulltime | -0.0468 | 0.0646 | 0.0337 | 0.0342 | 0.0416 | 0.0429 |
| | (-0.178 - 0.0842) | (-0.0360 - 0.165) | (-0.0844 - 0.152) | (-0.0309 - 0.0993) | (-0.132 - 0.215) | (-0.0419 - 0.128) |
| job_satisfied | 0.161*** | 0.231*** | 0.170*** | 0.187*** | 0.132*** | 0.182*** |
| | (0.124 - 0.199) | (0.199 - 0.264) | (0.117 - 0.224) | (0.114 - 0.260) | (0.0890 - 0.175) | (0.111 - 0.253) |
| generation = 2, Gen Y (1981-1996) | 0.0844* | -0.0131 | 0.0787 | -0.0140 | 0.0419 | 0.0352 |
| | (-0.00396 - 0.173) | (-0.0610 - 0.0348) | (-0.0182 - 0.176) | (-0.0602 - 0.0323) | (-0.0413 - 0.125) | (-0.0542 - 0.125) |
| generation = 3, Gen X (1965-1980) | 0.00260 | -0.0173 | -0.00447 | -0.0784* | -0.0734 | -0.00876 |
| | (-0.0839 - 0.0891) | (-0.0883 - 0.0538) | (-0.106 - 0.0967) | (-0.170 - 0.0133) | (-0.172 - 0.0255) | (-0.0967 - 0.0792) |
| generation = 4, Baby Boomers (1964 & Older) | -0.0290 | -0.0544* | -0.0172 | 0.0236 | -0.0698 | -0.00226 |
| | (-0.125 - 0.0672) | (-0.117 - 0.00852) | (-0.187 - 0.153) | (-0.0639 - 0.111) | (-0.168 - 0.0282) | (-0.0902 - 0.0857) |
| manufacturing | -0.00628 | 0.0925** | 0.00173 | 0.0901*** | 0.121*** | 0.243*** |
| | (-0.0629 - 0.0503) | (0.0173 - 0.168) | (-0.0494 - 0.0528) | (0.0357 - 0.144) | (0.0692 - 0.172) | (0.169 - 0.317) |
| Observations | 1,729 | 1,188 | 1,729 | 1,188 | 1,729 | 1,188 |
| Company Size | YES | YES | YES | YES | YES | YES |
| Country | YES | YES | YES | YES | YES | YES |
| Obs | Male | Female | Male | Female | Male | Female |

ci in parentheses

*** p<0.01, ** p<0.05, * p<0.1

Gender-specific analyses further indicate heterogeneity. AI use is positively associated with all well-being dimensions for both men and women, but the magnitudes differ: women exhibit stronger associations with mental health, while men show more pronounced associations with physical health. These differences are consistent with variation in job roles, task environments, and patterns of AI integration across genders.

This study has some limitations that should be discussed. The findings should be interpreted as associations rather than causal effects, as the cross-sectional design does not permit causal inference. Additional limitations include reliance on self-reported well-being measures, potential selection bias between AI users and non-users, and limited generalizability beyond the manufacturing and finance/insurance sectors examined. The data were collected in early 2022 and may not fully capture the impacts of more recent generative AI developments. Moreover, AI use is measured as a binary indicator, which does not reflect differences in the type or intensity of AI use; future research could explore how more granular measures of AI adoption relate to worker well-being.

## 5. Conclusion

This study provides cross-country evidence on how workplace AI use relates to workers' perceived well-being across mental health, job enjoyment, and physical health and safety in seven OECD economies. Using worker-level data from the OECD AI Surveys of Employers and Workers, the analysis reveals a consistent and robust positive association between AI use and all three well-being dimensions. The results suggest that AI tools are generally perceived as enhancing both the emotional and physical aspects of workers' experiences.

The results also reveal meaningful heterogeneity across demographic groups. Generation Y shows the strongest and most comprehensive associations, with positive relationships across mental, physical, and job-related outcomes. Generation X also exhibits positive associations, particularly for mental health and job enjoyment, though with smaller magnitudes. In contrast, Generation Z—despite high levels of AI use—shows positive associations primarily for job enjoyment, while Baby Boomers exhibit more limited associations concentrated in mental health. Gender-specific analyses further indicate differentiated patterns: women report stronger associations with mental health, whereas men report relatively larger associations with physical health. Together, these findings suggest that the relationship between AI use and well-being varies systematically across workers with different career stages, roles, and patterns of technology engagement.

Taken together, the evidence indicates that while AI use is broadly associated with improved worker well-being, these associations are not evenly distributed. Recognizing which groups experience stronger or weaker associations is important for policymakers and organizations seeking to integrate AI in ways that support inclusive and positive workplace outcomes. Attention to workers' skills, tasks, and career stages may help ensure that the potential benefits of AI adoption are more widely shared.

Several limitations should be noted. The cross-sectional nature of the data precludes causal inference, and the analysis relies on self-reported well-being measures, which may be subject to reporting bias. There may also be selection differences between AI users and non-users, and the findings may not generalize beyond the manufacturing and finance/insurance sectors examined. In addition, AI use is measured as a binary indicator and does not capture variation in the type or intensity of AI applications. Finally, because the data were collected in early 2022, the results may not fully reflect the effects of more recent developments in generative AI.

Future research could extend this analysis by examining longitudinal impacts of AI adoption, distinguishing between different types of AI tools (including generative AI), and exploring organizational factors that moderate the relationship between AI use and worker well-being. Such work would deepen our understanding of the social and economic implications of AI in the workplace and help inform more inclusive technology governance in an era of rapid technological change.

**Declaration of generative AI and AI-assisted technologies in the writing process**

During the preparation of this work, the author used NotebookLM to help screen the literature and GPT 5.2 to improve the readability and language in the writing of the manuscript. After using these tools, the author reviewed and edited the content as needed and takes full responsibility for the content of the published article.

Appendix Table 1: Main Regression Results (Logit Model: Marginal Effects)

| VARIABLES | (1)<br>Improved Mental Health | (2)<br>Improved Mental Health | (3)<br>Improved Enjoyment | (4)<br>Improved Enjoyment | (5)<br>Improved Physical Health | (6)<br>Improved Physical Health |
|---|---|---|---|---|---|---|
| ai user | 0.124*** | 0.109*** | 0.194*** | 0.181*** | 0.0892*** | 0.0773*** |
| | (0.0718 - 0.176) | (0.0508 - 0.167) | (0.141 - 0.247) | (0.126 - 0.235) | (0.0497 - 0.129) | (0.0373 - 0.117) |
| bachelor or higher | 0.0979*** | 0.0755*** | 0.141*** | 0.121*** | 0.123*** | 0.105*** |
| | (0.0472 - 0.149) | (0.0234 - 0.128) | (0.0961 - 0.185) | (0.0824 - 0.159) | (0.0777 - 0.168) | (0.0662 - 0.144) |
| male | 0.0925*** | 0.0749** | 0.0533* | 0.0378 | 0.0896*** | 0.0751*** |
| | (0.0287 - 0.156) | (0.0130 - 0.137) | (-0.00750 - 0.114) | (-0.0236 - 0.0992) | (0.0419 - 0.137) | (0.0295 - 0.121) |
| fulltime | 0.0344 | 0.0244 | 0.0456 | 0.0361 | 0.0443 | 0.0369 |
| | (-0.0275 - 0.0962) | (-0.0353 - 0.0841) | (-0.0198 - 0.111) | (-0.0246 - 0.0968) | (-0.0206 - 0.109) | (-0.0279 - 0.102) |
| job_satisfied | | 0.190*** | | 0.179*** | | 0.154*** |
| | | (0.163 - 0.216) | | (0.142 - 0.216) | | (0.128 - 0.181) |
| generation = 2, Gen Y (1981-1996) | 0.0146 | 0.0331 | 0.0246 | 0.0420 | 0.0222 | 0.0366 |
| | (-0.0579 - 0.0870) | (-0.0320 - 0.0982) | (-0.0488 - 0.0981) | (-0.0247 - 0.109) | (-0.0554 - 0.0998) | (-0.0331 - 0.106) |
| generation = 3, Gen X (1965-1980) | -0.0450 | -0.0191 | -0.0622 | -0.0374 | -0.0713 | -0.0509 |
| | (-0.133 - 0.0432) | (-0.0978 - 0.0595) | (-0.150 - 0.0258) | (-0.123 - 0.0482) | (-0.161 - 0.0188) | (-0.136 - 0.0347) |
| generation = 4, Baby Boomers (1964 & Older) | -0.0660 | -0.0529 | -0.0177 | -0.00538 | -0.0586 | -0.0484 |
| | (-0.161 - 0.0287) | (-0.142 - 0.0357) | (-0.174 - 0.138) | (-0.163 - 0.153) | (-0.140 - 0.0233) | (-0.126 - 0.0293) |
| manufacturing | 0.0175 | 0.0319 | 0.0219 | 0.0345 | 0.156*** | 0.167*** |
| | (-0.0333 - 0.0682) | (-0.0198 - 0.0837) | (-0.0261 - 0.0698) | (-0.0107 - 0.0797) | (0.109 - 0.203) | (0.120 - 0.214) |
| Observations | 2,917 | 2,917 | 2,917 | 2,917 | 2,917 | 2,917 |
| Company Size | YES | YES | YES | YES | YES | YES |
| Country | YES | YES | YES | YES | YES | YES |
| Obs | All | All | All | All | All | All |

ci in parentheses

*** p<0.01, ** p<0.05, * p<0.1

Appendix Table 2: Regression Results by Generation (Logit Model: Marginal Effects)

| | (1) | (2) | (3) | (4) | (5) | (6) | (7) | (8) | (9) | (10) | (11) | (12) |
|---|---|---|---|---|---|---|---|---|---|---|---|---|
| VARIABLES | Improved Mental Health | Improved Mental Health | Improved Mental Health | Improved Mental Health | Improved Enjoyment | Improved Enjoyment | Improved Enjoyment | Improved Enjoyment | Improved Physical Health | Improved Physical Health | Improved Physical Health | Improved Physical Health |
| ai user | -0.0141 | 0.163*** | 0.0688* | 0.108** | 0.141*** | 0.189*** | 0.195*** | 0.112 | -0.00980 | 0.121*** | 0.0434 | 0.0550 |
| | (-0.140 - 0.111) | (0.0720 - 0.254) | (-0.000109 - 0.138) | (0.0129 - 0.203) | (0.0512 - 0.230) | (0.122 - 0.256) | (0.125 - 0.265) | (-0.0310 - 0.254) | (-0.146 - 0.126) | (0.0700 - 0.172) | (-0.0288 - 0.116) | (-0.0526 - 0.163) |
| bachelor or higher | 0.0535 | 0.0517** | 0.133*** | -0.0736 | 0.119*** | 0.126*** | 0.120*** | 0.0879 | 0.0326 | 0.0856*** | 0.161*** | 0.0219 |
| | (-0.0426 - 0.150) | (0.00776 - 0.0956) | (0.0656 - 0.201) | (-0.228 - 0.0812) | (0.0618 - 0.176) | (0.0840 - 0.169) | (0.0328 - 0.207) | (-0.0230 - 0.199) | (-0.0223 - 0.0874) | (0.0253 - 0.146) | (0.0963 - 0.227) | (-0.101 - 0.145) |
| male | 0.0316 | 0.109** | 0.0406 | 0.0733 | 0.0149 | 0.0684* | 0.0394 | -0.0381 | 0.141*** | 0.0994*** | 0.0195 | 0.0368 |
| | (-0.0721 - 0.135) | (0.0207 - 0.197) | (-0.0446 - 0.126) | (-0.0244 - 0.171) | (-0.0765 - 0.106) | (-0.00572 - 0.143) | (-0.0503 - 0.129) | (-0.129 - 0.0526) | (0.0503 - 0.232) | (0.0274 - 0.171) | (-0.0441 - 0.0832) | (-0.0322 - 0.106) |
| fulltime | -0.0168 | 0.00930 | 0.0847 | -0.0449 | -0.00836 | 0.0208 | 0.0906 | -0.0166 | 0.00207 | 0.0751** | 0.0567 | -0.152 |
| | (-0.153 - 0.120) | (-0.0751 - 0.0937) | (-0.0725 - 0.242) | (-0.213 - 0.123) | (-0.186 - 0.169) | (-0.0248 - 0.0664) | (-0.0693 - 0.251) | (-0.166 - 0.133) | (-0.150 - 0.154) | (0.00466 - 0.145) | (-0.0333 - 0.147) | (-0.349 - 0.0454) |
| job_satisfied | 0.0983* | 0.234*** | 0.162*** | 0.141** | 0.0406 | 0.200*** | 0.200*** | 0.164*** | 0.0451 | 0.172*** | 0.168*** | 0.110 |
| | (-0.00705 - 0.204) | (0.183 - 0.286) | (0.119 - 0.205) | (0.0231 - 0.259) | (-0.0533 - 0.134) | (0.164 - 0.235) | (0.150 - 0.251) | (0.0534 - 0.275) | (-0.0756 - 0.166) | (0.129 - 0.216) | (0.106 - 0.231) | (-0.0470 - 0.267) |
| manufacturing | -0.0544 | 0.00997 | 0.0808** | 0.127 | 0.0167 | 0.0217 | 0.0557 | 0.0473 | 0.0442 | 0.119*** | 0.254*** | 0.275*** |
| | (-0.164 - 0.0556) | (-0.0569 - 0.0768) | (0.00481 - 0.157) | (-0.0375 - 0.292) | (-0.0480 - 0.0814) | (-0.0234 - 0.0668) | (-0.0176 - 0.129) | (-0.117 - 0.212) | (-0.120 - 0.208) | (0.0514 - 0.187) | (0.213 - 0.294) | (0.118 - 0.432) |
| Observations | 348 | 1,408 | 952 | 205 | 348 | 1,408 | 952 | 209 | 348 | 1,408 | 952 | 205 |
| Company Size | YES | YES | YES | YES | YES | YES | YES | YES | YES | YES | YES | YES |
| Country | YES | YES | YES | YES | YES | YES | YES | YES | YES | YES | YES | YES |
| Obs | Gen Z | Gen Y | Gen X | Baby Boomers | Gen Z | Gen Y | Gen X | Baby Boomers | Gen Z | Gen Y | Gen X | Baby Boomers |

ci in parentheses

*** p<0.01, ** p<0.05, * p<0.1

Appendix Table 3: Regression Results by Gender (Logit Model: Marginal Effects)

| VARIABLES | (1)<br>Improved Mental Health | (2)<br>Improved Mental Health | (3)<br>Improved Enjoyment | (4)<br>Improved Enjoyment | (5)<br>Improved Physical Health | (6)<br>Improved Physical Health |
|---|---|---|---|---|---|---|
| ai user | 0.0898*** | 0.118*** | 0.174*** | 0.184*** | 0.0814*** | 0.0700* |
| | (0.0392 - 0.140) | (0.0297 - 0.207) | (0.0991 - 0.248) | (0.122 - 0.246) | (0.0241 - 0.139) | (-0.00520 - 0.145) |
| bachelor or higher | 0.0768** | 0.0699** | 0.112*** | 0.129*** | 0.119*** | 0.0752*** |
| | (0.0114 - 0.142) | (0.00874 - 0.131) | (0.0647 - 0.160) | (0.0688 - 0.188) | (0.0656 - 0.172) | (0.0300 - 0.120) |
| fulltime | -0.0477 | 0.0633 | 0.0336 | 0.0338 | 0.0439 | 0.0407 |
| | (-0.183 - 0.0873) | (-0.0386 - 0.165) | (-0.0809 - 0.148) | (-0.0305 - 0.0981) | (-0.135 - 0.223) | (-0.0460 - 0.127) |
| job_satisfied | 0.161*** | 0.228*** | 0.171*** | 0.187*** | 0.131*** | 0.182*** |
| | (0.123 - 0.199) | (0.197 - 0.260) | (0.116 - 0.226) | (0.114 - 0.260) | (0.0865 - 0.175) | (0.111 - 0.252) |
| generation = 2, Gen Y (1981-1996) | 0.0829* | -0.0142 | 0.0756 | -0.0163 | 0.0367 | 0.0344 |
| | (-0.00412 - 0.170) | (-0.0636 - 0.0351) | (-0.0242 - 0.175) | (-0.0625 - 0.0300) | (-0.0457 - 0.119) | (-0.0608 - 0.130) |
| generation = 3, Gen X (1965-1980) | 0.00173 | -0.0182 | -0.00652 | -0.0818* | -0.0785 | -0.0100 |
| | (-0.0835 - 0.0869) | (-0.0896 - 0.0533) | (-0.109 - 0.0964) | (-0.173 - 0.00927) | (-0.174 - 0.0171) | (-0.103 - 0.0831) |
| generation = 4, Baby Boomers (1964 & Older) | -0.0308 | -0.0555* | -0.0193 | 0.0238 | -0.0742 | -0.00153 |
| | (-0.128 - 0.0668) | (-0.118 - 0.00690) | (-0.196 - 0.158) | (-0.0645 - 0.112) | (-0.172 - 0.0238) | (-0.0919 - 0.0888) |
| manufacturing | -0.00611 | 0.0930** | 0.00152 | 0.0907*** | 0.122*** | 0.242*** |
| | (-0.0624 - 0.0502) | (0.0190 - 0.167) | (-0.0516 - 0.0546) | (0.0378 - 0.144) | (0.0714 - 0.173) | (0.169 - 0.316) |
| Observations | 1,729 | 1,188 | 1,729 | 1,188 | 1,729 | 1,188 |
| Company Size | YES | YES | YES | YES | YES | YES |
| Country | YES | YES | YES | YES | YES | YES |
| Obs | Male | Female | Male | Female | Male | Female |

ci in parentheses

*** p<0.01, ** p<0.05, * p<0.1